\documentclass[a4paper,nofootinbib,twoside]{revtex4-2} 
\usepackage[left=20mm,right=20mm,top=25mm]{geometry} 
\usepackage[bookmarks=false,colorlinks]{hyperref}
\usepackage[english]{babel}
\usepackage[utf8]{inputenc}
\usepackage[T1]{fontenc}
\usepackage{amsmath,amsthm,amsfonts,amssymb,mathtools,bm,latexsym,stmaryrd}
\usepackage[table,dvipsnames]{xcolor}
\usepackage[pdftex]{graphicx}
\usepackage{adjustbox}
\usepackage{placeins}
\usepackage{lipsum}
\usepackage{csquotes}
\usepackage{comment}
\usepackage{subcaption}
\usepackage[bookmarks=false,colorlinks]{hyperref}
\hypersetup{urlcolor=NavyBlue, citecolor=NavyBlue}
\usepackage{ragged2e}
\usepackage{float}
\makeatletter
\let\newfloat\newfloat@ltx
\makeatother
\usepackage{algorithm}
\usepackage{algpseudocode}
\usepackage{braket}
\usepackage{bbm}

\DeclareCaptionJustification{justified}{\justifying}

\newcommand{\BLUE}[1]{\textcolor{black}{#1}}

\begin{document}

\title{Applying Grover-mixer quantum alternating operator ansatz algorithm to higher-order unconstrained binary optimization problems}

\author{Evgeniy O. Kiktenko}\email[Email address:~]{evgeniy.kiktenko@gmail.com}
\affiliation{National University of Science and Technology ``MISIS'', Moscow 119049, Russia}
\author{Elizaveta V. Krendeleva}
\affiliation{National University of Science and Technology ``MISIS'', Moscow 119049, Russia}
\author{Aleksey K. Fedorov}
\affiliation{National University of Science and Technology ``MISIS'', Moscow 119049, Russia}


\begin{abstract}
The quantum approximate optimization algorithm (QAOA) is among the leading candidates for achieving quantum advantage on near-term processors. While typically implemented with a transverse-field mixer (XM-QAOA), the Grover-mixer variant (GM-QAOA) offers a compelling alternative due to its global search capabilities. This work investigates the application of GM-QAOA to higher-order unconstrained binary optimization (HUBO) problems, also known as polynomial unconstrained binary optimization (PUBO), which form a general class of combinatorial optimization problems involving multivariable interactions. We present a comprehensive numerical study demonstrating that GM-QAOA, unlike XM-QAOA, exhibits monotonic improvement in performance with circuit depth and achieves superior results for HUBO problems within \BLUE{a layerwise optimization framework}. An important component of our approach is an analytical framework for modeling GM-QAOA dynamics, which enables a classical approximation of the optimal parameters and helps reduce the optimization overhead. Our resource-efficient parametrized version of GM-QAOA nearly matches the performance of \BLUE{the version optimized using the layerwise approach} while being significantly less demanding, making it a highly effective approach for complex optimization tasks. These findings highlight the potential of GM-QAOA and provide a practical pathway for its implementation on current quantum hardware.
\end{abstract}

\keywords{quantum computing, quantum algorithms, QAOA, Grover, HUBO}

\maketitle

\section{Introduction}\label{sec:Introduction}

Quantum computing has emerged as a powerful paradigm for addressing computational challenges that are intractable for classical systems. In the noisy intermediate-scale quantum (NISQ) era, variational quantum algorithms (VQAs) have become a cornerstone for practical applications~\cite{bharti2022noisy, cerezo2021variational}, demonstrating promise in areas such as quantum machine learning ~\cite{biamonte2017quantum}, quantum chemistry ~\cite{mcardle2020quantum}, quantum simulation ~\cite{yuan2019theory}, and linear algebra ~\cite{xu2021variational}. By harnessing the principles of quantum mechanics, such as superposition and entanglement, quantum processors offer the potential for substantial speedups in these and other domains, including cryptography, material science, and optimization ~\cite{fedorov2022quantum}.

Among the most promising approaches for near-term quantum devices is the quantum approximate optimization algorithm (QAOA) \cite{farhi2014quantum}, which has been theoretically shown to hold potential for quantum advantage ~\cite{farhi2016quantum} and has been successfully implemented on various quantum hardware platforms ~\cite{pagano2020quantum, harrigan2021quantum, zhou2020quantum}. In the literature, the acronym ``QAOA'' appears in two closely related forms: \emph{quantum approximate optimization algorithm}, emphasizing its goal-oriented nature, and \emph{quantum alternating operator ansatz}, highlighting the alternating structure of cost and mixer unitaries. The latter viewpoint has become increasingly relevant as generalized, problem-aware mixers have been developed to extend QAOA beyond the standard transverse-field construction.

\begin{figure}[H]
\centering
\includegraphics[width=0.9\textwidth]{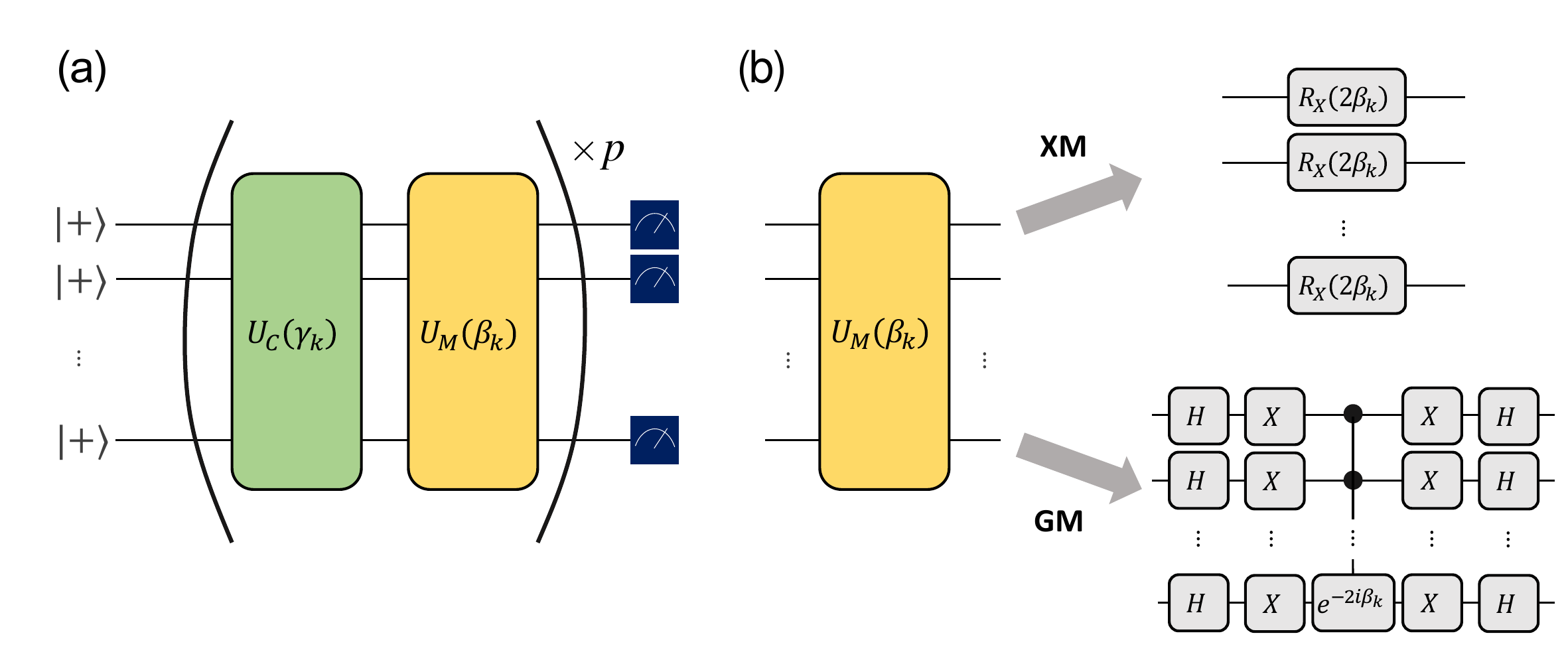}
\caption{(a) General structure of the QAOA circuit. (b) Two implementations of the mixing operator considered in this work: the standard transverse-field mixer (leading to XM-QAOA) and the Grover-type mixer (leading to GM-QAOA). Standard notation is used for the Hadamard gate, rotations about the Bloch $x$ axis, the Pauli-$X$ (inversion) gate, and the multicontrolled phase gate.}
\label{fig:scheme}
\end{figure}

A prominent example of such generalization is the Grover-mixer QAOA (GM-QAOA), which replaces the transverse-field mixer with a Grover-style diffusion operator~\cite{grover1996fast,bartschi2020grover, bridi2024analytical, xie2025performance,
golden2021threshold, benchasattabuse2023lower, pelofske2025biased,
ng2024analytical} (see also Fig.~\ref{fig:scheme}). The conceptual foundation for using Grover-based amplitude amplification in quantum simulation and optimization was previously demonstrated in the context of finding low-energy states of disordered Ising models~\cite{zhukov2025grover}, highlighting its potential beyond unstructured search. Unlike the standard mixer, which acts locally and updates qubits independently, the Grover mixer performs a global transformation that collectively rotates amplitudes across the entire computational basis. This global mixing dynamically couples all basis states, enabling more effective exploration of complex energy landscapes and potentially accelerating convergence. 

A growing body of work has examined the properties and limitations of Grover-type mixers. Analytical studies indicate that GM-QAOA performance is largely influenced by the statistical structure of the cost landscape, and that Grover mixers can offer at most quadratic sampling advantages, while typically requiring exponentially increasing depth to maintain a fixed approximation performance~\cite{bridi2024analytical,xie2025performance}. Extensions such as
threshold-based mixers have been proposed to improve parameter selection and empirical behavior in specific regimes~\cite{golden2021threshold}, while numerical investigations have highlighted distinctive features including significantly more uniform sampling across degenerate optima compared to transverse-field QAOA (XM-QAOA)~\cite{pelofske2025biased}. More recent analytical treatments have clarified the inherently nonlocal structure of multilayer Grover mixers and their connections to higher-order interactions in the
underlying hypergraph~\cite{ng2024analytical}. \BLUE{Complementary lines of research have further expanded the understanding of Grover-based approaches: from quantum algorithms for counting weighted ground states in spin Hamiltonians ~\cite{sundar2019quantum} to experimental realizations of advanced mixers on trapped-ion platforms ~\cite{zhu2023multi}, and most recently, the demonstration of quadratic speedup and fair-sampling properties of Grover-QAOA for constraint satisfaction problems such as 3-SAT ~\cite{zhang2025grover}.
At the same time, studies of QAOA applied to 2-SAT and 3-SAT problems have shown that its performance is highly sensitive to instance structure and density, with intrinsic limitations at fixed circuit depth~\cite{akshay2020reachability}. Systematic comparisons further indicate that the relative performance of local and global mixers depends strongly on problem size~\cite{golden2023quantum}. Complementary results for constrained optimization problems further show that appropriately tailored mixers are able to exploit problem structure and achieve improved performance~\cite{golden2023numerical}.
} In parallel, Lie-algebraic analysis has demonstrated that GM-QAOA possesses unusually rich dynamical expressivity: its dynamical Lie algebra is maximal among QAOA variants initialized with the same state, yielding explicit characterizations of conserved quantities and even provable avoidance of barren plateaus for broad classes of objective functions~\cite{tsvelikhovskiy2025provable}. Taken together, these results provide a comprehensive and nuanced picture of both the strengths and the intrinsic limitations of Grover-type mixers in variational quantum optimization.

A commonly noted limitation of the Grover mixer is its intrinsic nonlocality: implementing the corresponding diffusion operator generally requires highly entangling multiqubit operations, in stark contrast to the transverse-field mixer composed of simple single-qubit rotations. Such nonlocal structure makes the Grover mixer reminiscent of complex multicontrolled gates, including generalizations of the Toffoli gate. Nevertheless, recent theoretical~\cite{kiktenko2020scalable, nikolaeva2022decomposing,kiktenko2025colloquium, nikolaeva2024efficient} and experimental~\cite{nikolaeva2025scalable, chu2023scalable} progress in qudit-based architectures demonstrates that multiqubit entangling operations can be substantially simplified when
higher-dimensional degrees of freedom are available. 
\BLUE{In particular, qudits provide access to additional internal levels that can be temporarily populated during a computation and used as auxiliary storage for quantum information. In the context of decomposing multicontrolled operations such as Toffoli gates, these extra levels effectively replace the need for multiple ancillary qubits: intermediate logical conditions can be coherently accumulated within a single physical system by transferring population between its levels.
This mechanism enables significantly more resource-efficient implementations of multicontrolled gates, reducing both circuit depth and the number of required entangling operations compared to standard qubit-based decompositions. As a result, the hardware overhead associated with Grover-type mixers may be substantially alleviated on emerging multilevel quantum platforms, where such native high-dimensional encodings are naturally available.}

\BLUE{
This naturally raises the question of which optimization problems benefit most from globally acting mixing operations. A promising class is given by problems with inherently nonlocal objective functions. Higher-order unconstrained binary optimization (HUBO), also known as polynomial unconstrained binary optimization (PUBO), provides a broad and practically relevant framework, as it includes interactions among more than two binary variables. This higher-order structure goes beyond pairwise QUBO couplings and aligns naturally with mixing strategies that explore the configuration space collectively. HUBO problems have attracted attention not only in quantum computing but also in alternative hardware paradigms such as optical-inspired analog simulators~\cite{chermoshentsev2021polynomial}. Although any HUBO can be reduced to a quadratic unconstrained binary optimization (QUBO) problem via auxiliary variables~\cite{rosenberg1975reduction, boros2014quadratization, semenov2025technique, dattani2019quadratization}, this transformation introduces significant overhead, increasing both problem size and landscape complexity. Working directly with the native HUBO formulation is therefore appealing for quantum algorithms capable of handling higher-order interactions. Such interactions arise naturally in diverse domains, including machine learning~\cite{sejnowski1986higher, kolda2009tensor, chang2011libsvm}, computational biology~\cite{wei2014detecting, klein2008large}, and logistics and scheduling~\cite{ribeiro2008hybrid, bierwirth1999production}. Notable examples include the low-autocorrelation binary sequence (LABS) problem~\cite{shaydulin2024evidence}, portfolio optimization with higher-order moments~\cite{uotila2025higher}, and the vehicle routing problem with distance balancing~\cite{ikeuchi2025evaluating}. Compared to QUBO, HUBO formulations more faithfully capture complex dependencies but at the cost of increased combinatorial complexity, as both the number of terms and the ruggedness of the energy landscape grow with interaction order. This makes HUBO a natural and practically relevant setting for assessing the performance of quantum optimization strategies, particularly those employing global mixing mechanisms such as the Grover mixer.
}

In this work, we perform a detailed investigation of Grover-mixer QAOA applied to representative HUBO problem classes. Our comprehensive numerical study shows that GM-QAOA, unlike standard XM-QAOA, exhibits monotonic performance improvement with circuit depth and achieves superior results for models with high-order interactions. An additional contribution of this work is the development of an analytical framework for modeling the dynamics of GM-QAOA layers, enabling accurate classical approximation of near-optimal parameters and reducing the overhead of variational optimization. We demonstrate that this resource-efficient parametrization achieves performance close to that of \BLUE{layerwise} optimized GM-QAOA while requiring significantly fewer circuit evaluations, making it far less demanding experimentally. Together, these results highlight the potential of Grover-type global mixers for complex high-order optimization landscapes and provide a practical pathway for near-term implementations. \BLUE{In contrast to prior studies on predominantly low-order problems, we focus on intrinsically higher-order interactions, where global mixers exhibit qualitatively different behavior.}

This paper is organized as follows. In Sec.~\ref{sec:preliminaries}, we provide the necessary preliminaries on HUBO problems and the GM-QAOA framework. Section~\ref{sec:comparison} presents a comparative analysis of GM-QAOA and standard XM-QAOA on a set of HUBO instances. Section~\ref{sec:analytics} is dedicated to our analytical contribution: we develop a mathematical model of the GM-QAOA dynamics and present a resource-efficient strategy for approximating optimal algorithm parameters. In Sec.~\ref{sec:perf}, we analyze the performance of GM-QAOA using analytically preoptimized parameters, and benchmark this approach against constant-parameter and \BLUE{layerwise} optimized QAOA variants across different problem sizes and Hamiltonian localities. Finally, we conclude in Sec.~\ref{sec:concl} with a summary of our findings and an outlook on future research directions.

\section{Preliminaries}\label{sec:preliminaries}

\subsection{HUBO problems}

HUBO represents a significant generalization of the well-studied QUBO framework. While QUBO problems are restricted to pairwise interactions between binary variables, HUBO problems capture complex multivariable correlations through higher-degree polynomial terms. This extension enables more accurate modeling of real-world optimization challenges where simultaneous interactions among multiple entities play a crucial role.

Formally, a HUBO (PUBO) problem is defined as the minimization of a real-valued pseudo-Boolean function. The cost function $E:\{\pm 1\}^n \rightarrow \mathbb{R}$ can be written in the spin representation as
\begin{equation}
    E(\mathbf{s}) = \sum_{d=1}^D \sum_{i_1<\cdots<i_d} J_{i_1\ldots i_d} \, s_{i_1}\cdots s_{i_d},
\end{equation}
where $\mathbf{s}=(s_1,\ldots,s_n)$ is a configuration of spin variables $s_i\in\{\pm 1\}$, $J_{i_1\ldots i_d}$ are real interaction coefficients, and $D$ is a HUBO \BLUE{order} parameter\BLUE{, where $D=2$ corresponds only to pairwise interactions, and $D>2$ includes all connections between themselves up to the order $D$}. Throughout this work, we denote the objective by $E(\mathbf{s})$ to emphasize its interpretation as an energy function, since the optimization task corresponds to identifying its ground-state configurations.

In the context of mapping to a quantum Hamiltonian, the Ising formulation is often more direct. The transformation follows the standard approach of replacing classical spin variables with quantum operators: $s_{i} \mapsto Z_i$, where $Z_i$ denotes the Pauli-$Z$ operator acting on the $i^{\rm th}$ qubit. This mapping preserves the eigenvalue structure, with computational basis states $\ket{z}$ corresponding to classical configurations \textbf{s} (hereinafter, we use the correspondence $0 \leftrightarrow +1$ and $1 \leftrightarrow -1$ between bit and sign variables). 
The resulting quantum Hamiltonian is constructed as
\begin{equation} \label{eq:cost_Hamiltonian}
    H_{C}=\sum_{d=1}^D\sum_{i_1< \ldots<i_d}J_{i_1\ldots i_d} Z_{i_1}\otimes Z_{i_2}\ldots\otimes Z_{i_d}
\end{equation}
This $D$-local Hamiltonian \cite{kempe2006complexity} contains multiqubit interaction terms that directly encode the higher-order correlations present in the original optimization problem. The ground state of $H_C$
corresponds to the optimal solution, with the ground energy equaling the minimal value of the cost function.

In analyzing HUBO problem complexity, it is essential to consider not only the cardinality of interaction terms but also the topological structure of the underlying hypergraph. Properties such as sparse connectivity, bounded hyperedge size, or specific graph minors can dramatically reduce the effective complexity.

Quantitatively, for $D=2$ we obtain the QUBO formulation with $O(n^2)$ terms in the worst case, though sparse instances may exhibit only $O(n)$ complexity. For arbitrary order $D$, the theoretical maximum grows combinatorially as $\sum_{d=1}^D \binom{n}{d}$, creating substantial challenges for classical optimization -- particularly when $D>2$ -- despite potential alleviation through structural sparsity~\cite{shaydulin2024evidence}.

\subsection{XM- and GM-QAOA}

QAOA is a hybrid quantum-classical algorithm designed to find approximate solutions to combinatorial optimization problems. The algorithm prepares a parametrized quantum state through the iterative application of two unitary operators derived from problem-specific and mixer Hamiltonians.

The quantum state generated by a $p$-layer QAOA circuit [shown in Fig.~\ref{fig:scheme}(a)] is given by the unitary
sequence
\begin{equation}
    \ket{\psi(\boldsymbol{\beta},\boldsymbol{\gamma})}
    = U_M(\beta_p)U_C(\gamma_p)\cdots U_M(\beta_1)U_C(\gamma_1)\ket{\psi_0},\quad 
    U_M(\beta)=e^{-i\beta H_M},\quad U_C(\gamma)= e^{-i\gamma H_C}.
\end{equation}
Here, $H_C$ is the cost Hamiltonian encoding the objective function of the
optimization problem, such that its ground state corresponds to the optimal
solution. The mixer Hamiltonian $H_M$ does not commute with $H_C$ and induces
transitions between its eigenstates. The vectors
$\boldsymbol{\beta}=(\beta_1,\ldots,\beta_p)$ and
$\boldsymbol{\gamma}=(\gamma_1,\ldots,\gamma_p)$ are the $2p$ variational
parameters optimized during the algorithm. The initial state $\ket{\psi_0}$ is
chosen to be an energy-extremal eigenstate of $H_M$.

The mixer Hamiltonian is crucial for exploring the feasible space of the problem. The choice of 
$H_M$ defines the specific QAOA variant and significantly impacts the algorithm's performance and the reachability of the optimal solution [see Fig.~\ref{fig:scheme}(b)].

The most common choice, inspired by quantum annealing, is the transverse-field mixer:
\begin{equation}
    H_{X}=\sum_{i=1}^nX_i
\end{equation}
where $X_j$ is the Pauli-$X$ operator on qubit $j$. The resulting unitary 
\begin{equation}
    U_X(\beta)=\prod_{j=1}^n e^{-i\beta X_j}
\end{equation}
implements simultaneous local $X$ rotations on all qubits. This mixer induces transitions between neighboring states in the computation basis by flipping individual bits, facilitating a local search through the solution space.

An advanced alternative is the Grover-type mixer, which promotes a more global search. It is defined using the projection onto the uniform superposition state $\ket{\rm sym}=\frac{1}{\sqrt{2^n}}\sum_{\bf s}{\ket{{\bf s}}}$:
\begin{equation}
    H_{G}=2\ket{\rm sym}\bra{\rm sym}.
\end{equation}
The corresponding unitary evolution operator has the form
\begin{equation}
    U_G(\beta)=e^{-i\beta H_G}= I + (e^{-i2\beta} - 1)|\rm sym\rangle\langle\rm sym|,
\end{equation}
where $I$ stands for the idenitity matrix.
This operator applies a conditional phase shift only to the $\ket{\rm sym}$ component, effectively performing a selective, global rotation in the Hilbert space. Unlike the local rotations of $U_X$, $U_G$ can create superpositions and interference patterns that simultaneously involve all possible states.

The fundamental difference between the two mixers lies in their interaction with the cost Hamiltonian. The sequence of $U_X(\beta_k)$ and $U_C(\gamma_k)$ can be interpreted as a discretized control sequence steering the system in Hilbert space. The local nature of $H_X$ means that the algorithm explores the solution space through a series of local moves, which can be advantageous for problems with a benign, local structure but may lead to slower convergence for problems with complex, long-range correlations. The GM-QAOA ansatz leverages global operations from the outset. The mixer $U_G(\beta_k)$ is equivalent to the diffusion operator in Grover's algorithm. When combined with the phase oracle $U_C(\gamma_k)$, it can amplify the probability amplitude of low-energy states more efficiently for certain problem classes. This can lead to a faster convergence (in terms of the number of layers 
$p$) and a higher approximation ratio at fixed 
$p$, particularly for problems where the optimal solution is markedly different from the average.

\subsection{Layerwised optimization}~\label{sec:layer-wised optimization}

A critical aspect of successfully implementing the QAOA is the efficient optimization of the parameter vectors $\bf \beta$ and $\bf \gamma$. As the number of layers $p$ increases, the optimization landscape becomes more complex, and the task of finding optimal parameters becomes computationally challenging. To address this, a layerwise optimization strategy is often employed.

The core idea of this strategy is to optimize the parameters incrementally. Instead of optimizing all 
$2p$ parameters simultaneously for a depth-$p$ circuit, one starts with a single-layer circuit and finds optimal parameters ($\beta_1, \gamma_1$). These parameters are then held fixed, and a new layer is added. The optimization for depth  $i+1$ then proceeds by optimizing only the new parameters ($\beta_{i+1}, \gamma_{i+1}$), initializing them based on the values from the previous layer or a heuristic strategy.

In this work, we use the probability of measuring the ground-state energy,
$P(E_{\min})$, as the objective function for optimizing the variational
parameters. For a given energy value $E$, this probability is defined as
\begin{equation}
    P(E)=\sum_{\mathbf{s}: \, E(\mathbf{s})=E}
    \bigl|\braket{\mathbf{s}|\psi(\boldsymbol{\beta},\boldsymbol{\gamma})}\bigr|^{2},
\end{equation}
where $\ket{\mathbf{s}}$ denotes the computational basis state corresponding to
the spin configuration $\mathbf{s}$, and $E_{\min}=\min_{\mathbf{s}} E(\mathbf{s})$
is the minimum value of the cost function. Thus, $P(E_{\min})$ represents the
total probability weight that the QAOA output state assigns to all configurations
achieving the optimal cost value.

\BLUE{It is important to note that while this layerwise optimization strategy is computationally efficient and has been employed in prior QAOA studies, it constitutes a heuristic approach. By fixing previously optimized parameters and only varying the angles of the newest layer, we restrict the search space and may converge to parameters that are sub-optimal compared to a full re-optimization of all $2p$ angles simultaneously. This limitation is particularly relevant for the transverse-field mixer, where such layerwise fixing has been observed to lead to a performance plateau. Therefore, the comparative results presented in this work should be interpreted within the context of this specific, resource-efficient optimization protocol.}

\BLUE{In this work, we consider two approaches to obtaining the layerwise optimized parameters. The first, which we refer to as standard GM-QAOA, employs numerical optimization (via quantum circuit simulation) to find the parameters \{$\beta_k, \gamma_k$\} that maximize $P(E_{\rm min})$. The second, introduced in Sec.~\ref{sec:analytics} and denoted as GM-QAOA(a), uses an analytical approximation to preoptimize the parameters classically, without recourse to quantum simulation.}

\section{Comparison on HUBO problems} \label{sec:comparison}

\BLUE{
In this section, we present a systematic comparison between the XM-QAOA and GM-QAOA frameworks applied to two representative combinatorial problems: a Max-Cut-like problem on random hypergraphs and the Sherrington-Kirkpatrick (SK) spin-glass model. Both problems are formulated as HUBO instances with varying locality and problem size.}

\BLUE{
For the SK model with maximum order $D$, all coupling coefficients $J_{i_1 \ldots i_d}$ for $d = 2,\ldots,D$ and all index combinations $i_1 < \ldots < i_d$ are drawn independently from the standard normal distribution, $J_{i_1 \ldots i_d} \sim \mathcal{N}(0,1)$ (i.e., with zero mean and unit variance), while linear terms are set to zero. This construction includes all possible interactions up to order $D$. For example, for $n = 14$ and $D = 4$, the instance contains $\binom{14}{2} + \binom{14}{3} + \binom{14}{4} = 1456$ interaction terms.
For the Max-Cut-like (hereafter Max-Cut) hypergraph instances, we adopt a Bernoulli generation scheme: each possible hyperedge of size $d$ ($2 \leq d \leq D$) is included independently with probability $1/2$, with corresponding coupling coefficients $J_{i_1 \ldots i_d} \in \{0,1\}$. This results in a dense random hypergraph. For the same parameters $n = 14$ and $D = 4$, the expected number of terms is approximately 728.}
For each problem configuration, 100 random instances were generated and subsequently solved \BLUE{using the layerwise optimization approach. The full dataset is available in Ref.~\cite{supplementary_data}.}

\begin{figure}
    \centering
    \includegraphics[width=\textwidth]{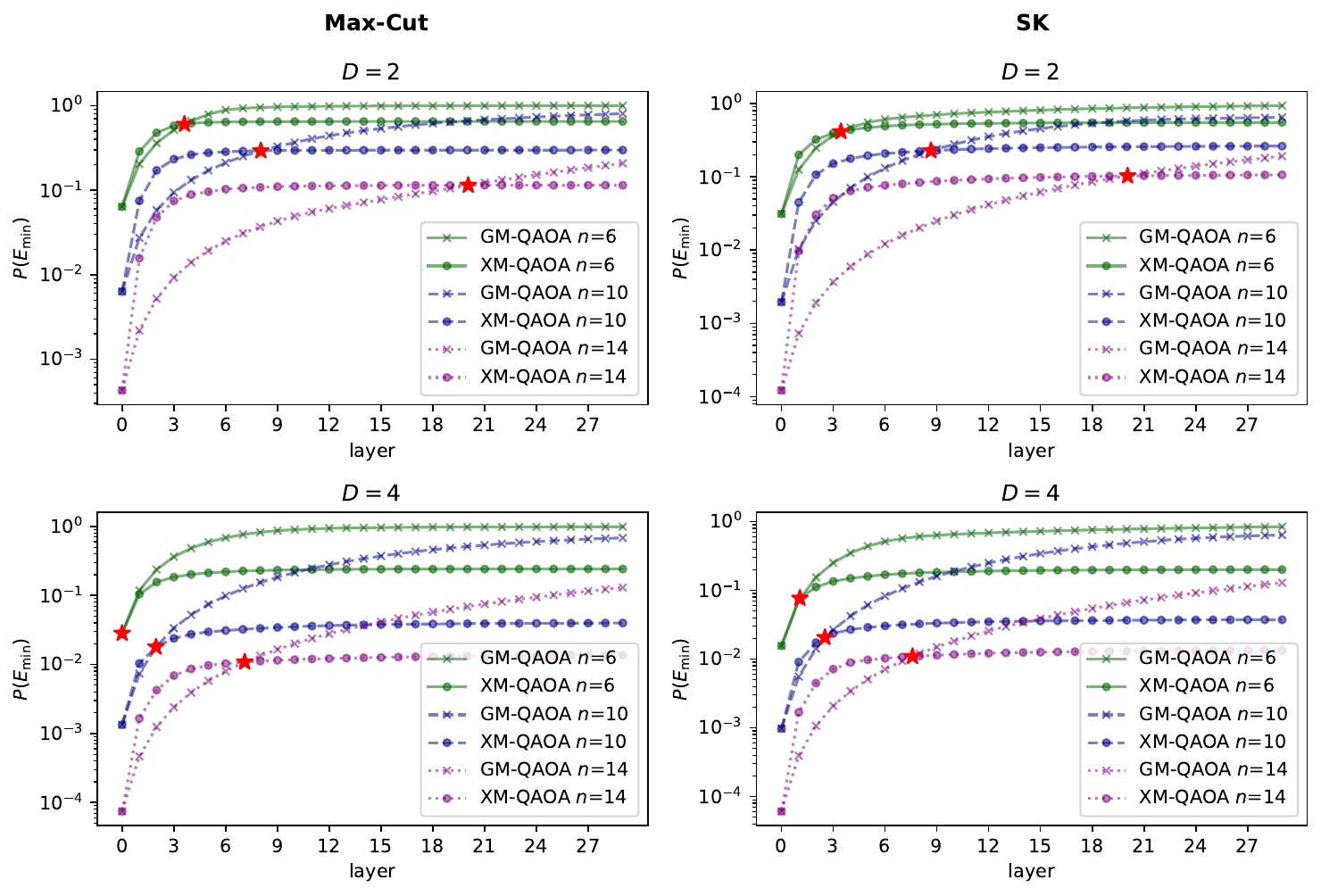}
    \caption{Performance comparison between GM-QAOA and XM-QAOA for the Max-Cut-like problem on random hypergraphs (left panels) and the SK-type spin-glass model (right panels). The plots show the ground-state success probability $P(E_{\min})$ as a function of the circuit depth (layer number) for different system sizes $n \in \{6,10,14\}$ and interaction orders $D=2$ (top row) and $D=4$ (bottom row). \BLUE{Here, $D$ denotes the maximum locality of the cost Hamiltonian: $D=2$ corresponds to standard quadratic Ising models (i.e., conventional Max-Cut and SK), while $D>2$ represents their higher-order (hypergraph) generalizations including interactions up to order $D$.}
    Red stars indicate the minimal circuit depth (critical point) at which GM-QAOA first surpasses the corresponding XM-QAOA performance for each instance family. Each data point represents an average over 100 randomly generated problem instances\BLUE{, optimized using the layerwise protocol from Sec.~\ref{sec:layer-wised optimization}}
    }
    \label{fig: comp_prob_algo_mc_sk}
\end{figure}

Figure~\ref{fig: comp_prob_algo_mc_sk} illustrates the averaged results for $D=\{2, 4\}$ and $n=\{6,10,14\}$. The XM-QAOA curves exhibit rapid initial growth in $P(E_{\rm min})$ at low circuit depths, but quickly saturate, forming a plateau. This indicates that increasing the number of layers beyond a certain point does not improve solution quality. In contrast, GM-QAOA displays a markedly different behavior: it exhibits a slower, monotonic increase in performance, gradually rising without saturation within the tested range. Notably, the GM-QAOA curve eventually intersects the XM-QAOA plateau at a critical depth (marked with red stars), beyond which it consistently outperforms the transverse-field approach.

Two key trends emerge from the results. First, as the system size $n$ increases, the absolute values of $P(E_{\rm min})$ decrease for both algorithms, and the critical depth where GM-QAOA surpasses XM-QAOA shifts to larger values. This implies that GM-QAOA requires deeper circuits to outperform XM-QAOA as the problem scale grows.

More importantly, we observe a fundamental difference in how each algorithm responds to increased Hamiltonian locality $D$. For $D=2$, XM-QAOA achieves relatively high success probabilities, whereas for $D>2$, its performance drops significantly and remains consistently low. In stark contrast, GM-QAOA maintains nearly constant performance across different values of $D$. This resilience to higher-order interactions constitutes the principal advantage of the Grover-style mixer. Specifically, for $D=4$, the critical depth at which GM-QAOA exceeds the XM-QAOA plateau is approximately three times smaller than for $D=2$, highlighting its efficiency in tackling high-order optimization problems.

\begin{figure}
    \centering
    \includegraphics[width=\textwidth]{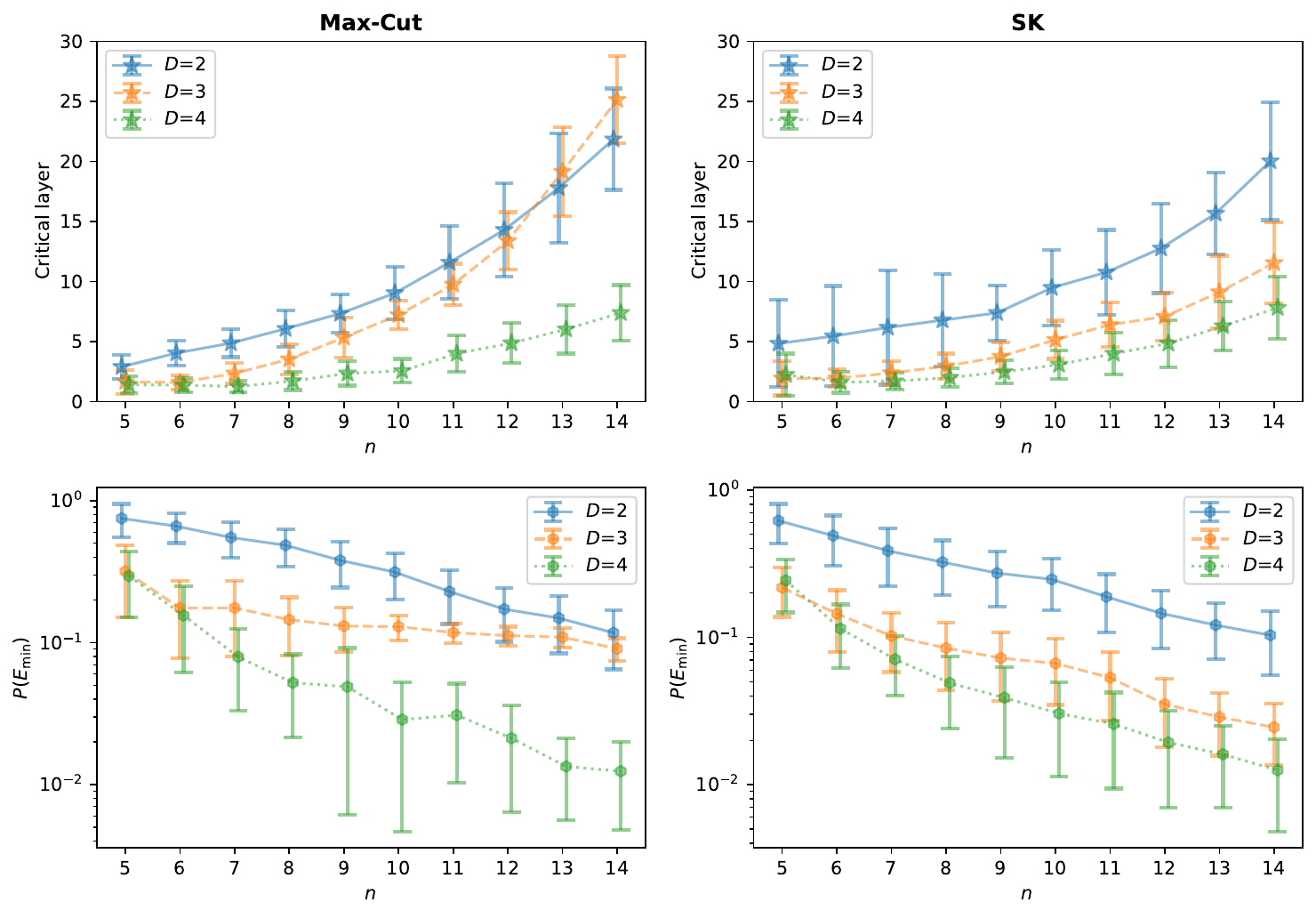}
    \caption{Changing the critical depth, defined as the minimum circuit depth at which GM-QAOA outperforms XM-QAOA, and the corresponding success probability as functions of the problem parameters. Error bars denote the standard deviation over 100 random problem instances.}
    \label{fig:phase_graphs_cr_point_mc_sk}
\end{figure}

Figure~\ref{fig:phase_graphs_cr_point_mc_sk} presents an analysis of the same problem instances used in Fig.~\ref{fig: comp_prob_algo_mc_sk} for two problem classes: Max-Cut (left panels) and the
SK model (right panels). The upper panels show the
dependence of the averaged critical layer index --- defined as the circuit depth beyond
which GM-QAOA begins to outperform XM-QAOA --- on the system size $n$ for different
interaction orders $D$. The lower panels display the corresponding success
probabilities evaluated at these critical depths.

These data reveal several key scaling properties. First, the critical depth exhibits monotonic growth with increasing system size $n$ for both problem types. This indicates that larger systems require deeper quantum circuits to achieve performance superiority of GM-QAOA over XM-QAOA. Notably, this growth is most pronounced for quadratic interactions ($D=2$), where the critical depth increases rapidly with 
$n$. In contrast, for higher-order interactions ($D=4$), the required depth remains substantially lower across all system sizes, highlighting the particular efficiency of GM-QAOA for high-degree optimization problems.

The success probability $P(E_{\rm min})$ at the critical point displays approximately exponential decay with increasing 
$n$. This behavior aligns with theoretical expectations for quantum optimization algorithms. Comparative analysis reveals that  $P(E_{\rm min})$ decays more rapidly for hypergraphs with a higher interaction degree, while systems with quadratic interactions maintain higher success probabilities across the studied range of $n$.

\section{Analytical approximation of GM-QAOA state dynamics}~\label{sec:analytics}
This section introduces a mathematical framework for modeling the layerwise evolution of GM‑QAOA amplitudes, enabling classical preoptimization of variational parameters. By adopting an energy‑resolved representation and assuming a Gaussian distribution of cost‑function values, we derive a compact recurrence for the state amplitudes. Combined with an extreme‑value‑theory (EVT) estimate of the spectral minimum, the model yields a resource‑efficient strategy to determine near‑optimal angles without costly quantum evaluations. 

\subsection{General framework and disorder-averaged dynamics}

Consider the quantum state at the $k^{\rm th}$ layer of GM-QAOA. It can be written as
\begin{equation}
    \ket{\boldsymbol{\beta}^{(k)},\boldsymbol{\gamma}^{(k)}} =
    \sum_{\mathbf{s}} \Psi_k(\mathbf{s}) \ket{\mathbf{s}},
\end{equation}
where $\boldsymbol{\beta}^{(k)} = (\beta_1,\ldots,\beta_k)$ and
$\boldsymbol{\gamma}^{(k)} = (\gamma_1,\ldots,\gamma_k)$ are truncated parameter
vectors (the total number of layers is $p$).
The coefficients $\Psi_k(\mathbf{s})$ denote probability amplitudes in the
computational basis. Although these amplitudes depend implicitly on
$\boldsymbol{\beta}^{(k)}$ and $\boldsymbol{\gamma}^{(k)}$, this dependence is
suppressed for notational simplicity.

The initial state corresponds to the uniform superposition, with amplitudes
$\Psi_0(\mathbf{s}) = 2^{-n/2}.$
Applying a single GM-QAOA layer, consisting of the problem (cost) unitary followed
by the Grover-type mixer, leads to the following update rule for the amplitudes:
\begin{equation}
    \Psi_k(\mathbf{s}) =
    \left(e^{-2 i \beta_k} - 1\right) \Theta_k
    + e^{-i \gamma_k E(\mathbf{s})} \Psi_{k-1}(\mathbf{s}),
\end{equation}
where
\begin{equation}
    \Theta_k =
    \frac{1}{2^n}
    \sum_{\mathbf{s}}
    e^{-i \gamma_k E(\mathbf{s})} \Psi_{k-1}(\mathbf{s})
\end{equation}
is the average amplitude over all computational basis states after applying the
problem-specific unitary.

Importantly, the evolution of the amplitudes $\Psi_k(\mathbf{s})$ depends on a basis
state $\mathbf{s}$ only through the corresponding value of the cost function
$E(\mathbf{s})$. This observation allows us to reparametrize the amplitudes in terms
of energy,
\begin{equation}
    \Psi_k(\mathbf{s}) \;\rightarrow\; \Psi_k(E),
\end{equation}
which forms the basis for the analytical approximation developed below.

We now treat the cost-function values $E(\mathbf{s})$ for different basis states
$\mathbf{s}$ as realizations of a random variable. For a random problem instance,
we approximate $E(\mathbf{s})$ as being drawn from a probability distribution
$f(E)$. Within this disorder-averaged description, the average amplitude
$\Theta_k$ can be expressed as an expectation value:
\begin{equation}
    \Theta_k \approx
    \int e^{-i \gamma_k \mathcal{E}} \,
    \Psi_{k-1}(\mathcal{E}) \,
    f(\mathcal{E}) \,
    \mathrm{d}\mathcal{E}
    \;\equiv\;
    \left\langle
        e^{-i \gamma_k \mathcal{E}} \Psi_{k-1}(\mathcal{E})
    \right\rangle_{\mathcal{E}} .
\end{equation}

As a result, we obtain the following recursive relation for the energy-resolved
probability amplitudes:
\begin{equation} \label{eq:recursive_expression}
    \Psi_k(E) =
    \left(e^{-2 i \beta_k} - 1\right)
    \left\langle
        e^{-i \gamma_k \mathcal{E}} \Psi_{k-1}(\mathcal{E})
    \right\rangle_{\mathcal{E}}
    + e^{-i \gamma_k E} \Psi_{k-1}(E).
\end{equation}

To determine the optimal GM-QAOA parameters
$\boldsymbol{\gamma}$ and $\boldsymbol{\beta}$, we maximize the probability of
obtaining a low-energy outcome,
\begin{equation}
    \left| \Psi_p\!\left(E_{\min}^{\mathrm{est}}\right) \right|^2
    \;\rightarrow\; \max ,
\end{equation}
where $E_{\min}^{\mathrm{est}}$ denotes an estimate of the location of the minimum
energy in the cost-function landscape.

\subsection{Gaussian energy distribution}

We now consider the specific case in which the energy distribution $f(E)$ is Gaussian with zero mean and variance $\sigma^2$:
\begin{equation}
    f(E) =
    \frac{1}{\sqrt{2\pi\sigma^2}}
    \exp\!\left(-\frac{E^2}{2\sigma^2}\right).
\end{equation}
The vanishing mean follows directly from the structure of the cost Hamiltonian
$H_C$, which is assumed to be composed of traceless operators
[see Eq.~\eqref{eq:cost_Hamiltonian}]. Indeed,
\begin{equation}
    \langle E(\mathbf{s}) \rangle_{\mathbf{s}}
    =
    \frac{1}{2^n}
    \sum_{\mathbf{s}} E(\mathbf{s})
    =
    \frac{1}{2^n} \mathrm{Tr}\, H_C
    =
    0.
\end{equation}

The variance $\sigma^2$ is determined by the second moment of the energy
distribution and can be expressed as
\begin{equation}
    \sigma^2
    =
    \langle E^2(\mathbf{s}) \rangle_{\mathbf{s}}
    =
    \frac{1}{2^n} \mathrm{Tr}\!\left(H_C^2\right)
    =
    \sum_{d=1}^D
    \sum_{i_1 < \ldots < i_d}
    J_{i_1 \ldots i_d}^2 ,
\end{equation}
where $D$ denotes the maximum interaction order of the cost Hamiltonian.

Substituting the Gaussian distribution into the recursive relation
Eq.~\eqref{eq:recursive_expression}, we find that the probability amplitude at
layer $k$ admits the following decomposition:
\begin{equation}
    \Psi_k(E) = A_k + B_k(E),
\end{equation}
where $A_k$ is an energy-independent contribution, while $B_k(E)$ captures the
explicit energy dependence.

For $k \geq 1$, these terms take the form
\begin{equation} \label{eq:Ak}
    A_k =
    \left(e^{-2 i \beta_k} - 1\right)
    \sum_{i=1}^{k}
    \exp\!\left[
        -\frac{\sigma^2}{2}
        \left(
            \gamma_k^2 + \gamma_{k-1}^2 + \ldots + \gamma_{k-i+1}^2
        \right)
    \right]
    A_{k-i},
\end{equation}
and
\begin{equation} \label{eq:Bk}
    B_k(E) =
    \sum_{j=1}^{k}
    A_{k-i}
    \exp\!\left[
        - i
        \left(
            \gamma_k + \gamma_{k-1} + \ldots + \gamma_{k-j+1}
        \right) E
    \right],
\end{equation}
with the initial conditions $A_0 = 2^{-n/2}$ and $B_0(E) = 0$.

\subsection{Minimum-energy approximation}

The recursive formulation of multilayer GM-QAOA dynamics requires an estimate of
the location of the minimum energy $E_{\rm min}$, which serves as a target value in
the optimization of the final-state probability distribution. Since the exact
ground-state energy is generally unknown, we employ tools from EVT to obtain a statistically motivated estimate based on an assumed distribution
of energy levels.

EVT provides a rigorous framework for characterizing the
statistics of extrema of large samples of random variables. According to the
Fisher-Tippett-Gnedenko theorem
\cite{fisher1928limiting, gnedenko1943distribution}, properly normalized extrema of
independent and identically distributed random variables converge to one of three
universal distributions, collectively described by the generalized extreme value
(GEV) distribution
\cite{jenkinson1955frequency, pickands1975statistical, coles2001introduction,
beirlant2006statistics, haan2006extreme},
\begin{equation}
    G(x) =
    \exp\!\left\{
        -\left[
            1 + \xi \left( \frac{x - \tilde\mu}{\tilde\sigma} \right)
        \right]^{-1/\xi}
    \right\},
    \quad
    1 + \xi \left( \frac{x - \tilde\mu}{\tilde\sigma} \right) > 0,
\end{equation}
where $\tilde\mu$ is the location parameter, $\tilde\sigma>0$ is the scale parameter, and $\xi$
is the shape parameter determining the distribution class. The cases $\xi=0$,
$\xi>0$, and $\xi<0$ correspond to the Gumbel, Fr\'echet, and Weibull distributions,
respectively.

Although EVT is conventionally formulated for maxima, the statistics of minima can
be obtained by applying the theory to the random variable $-E$. For energy levels
modeled as independent Gaussian random variables,
$E \sim \mathcal{N}(\mu,\sigma^2)$ (in our case $\mu=0$), the distribution belongs to
the Gumbel universality class, i.e., $\xi=0$.

The probability density function of the Gumbel distribution for the minimum energy
is then given by
\begin{equation}
    f_G(x) =
    \frac{1}{\beta_G}
    \exp\!\left(
        \frac{x - \mu_G}{\beta_G}
        -
        e^{\frac{x - \mu_G}{\beta_G}}
    \right),
\end{equation}
where the location and scale parameters are expressed in terms of order statistics
of $N$ independent Gaussian samples as
\begin{equation}
    \mu_G = \mu + \sigma \Phi^{-1}\!\left( \frac{1}{N} \right),
    \qquad
    \beta_G =
    \sigma \left[
        \Phi^{-1}\!\left( \frac{1}{N} \right)
        -
        \Phi^{-1}\!\left( \frac{1}{eN} \right)
    \right].
\end{equation}
Here $N = 2^n$ denotes the total number of computational basis states, and
$\Phi^{-1}$ is the quantile function of the standard normal distribution.

We use the mode of the Gumbel distribution as an estimate of the minimum energy,
which directly yields
\begin{equation}
    E_{\rm min}^{\rm est}
    =
    \mu_G
    =
    \sigma \Phi^{-1}\!\left( \frac{1}{2^n} \right).
\end{equation}
This estimate provides a simple and analytically tractable target energy for
GM-QAOA parameter optimization, while correctly capturing the exponential growth of
the effective search space with system size under the Gaussian energy approximation.

A closely related estimate was introduced in
Ref.~\cite{zhukov2025grover} for Ising problems with normally distributed
couplings. In that work, the authors also proposed an explicit large-$n$ approximation
to the Gaussian quantile, which leads to the expression
\begin{equation}
    E_{\rm min}^{\rm est}
    \approx
    -\sigma\sqrt{2\ln 2}\,\sqrt{n}
    \left(
        1 + \frac{1}{4\ln 2}\,\frac{\ln n}{n}
    \right)^{-1},
\end{equation}
reproducing the leading random-energy-model scaling
$E_{\rm min}\sim -\sigma\sqrt{2 n\ln 2}$. This closed-form approximation is
particularly convenient when numerical evaluation of $\Phi^{-1}$ is
impractical.

\section{Performance analysis} \label{sec:perf}

In this section, we present a resource-efficient strategy for selecting GM-QAOA
parameters that avoids extensive quantum hardware runtime. The approach builds
upon the analytical framework developed in the previous sections, combined with
the extreme-value-theory-based estimate of the minimum energy
$E_{\rm min}^{\rm est}$, and enables classical preoptimization of the variational
parameters $\{\beta_k,\gamma_k\}$.

The central idea is to maximize the analytically derived probability
$P(E_{\rm min}^{\rm est})$ of obtaining the minimum-energy configuration as a
function of the circuit parameters. 
This is achieved via a layer-by-layer numerical optimization of $P(E_{\rm min}^{\rm est})$, using the expressions for the two contributions to the probability amplitudes given in Eqs.~\eqref{eq:Ak} and \eqref{eq:Bk}.
The resulting optimized angles can then be
directly used in GM-QAOA circuits executed on quantum hardware, thereby reducing
or completely eliminating the need for iterative hybrid optimization. In what
follows, we refer to this analytically optimized variant as GM-QAOA(a).

We also compare our approach with the simplified GM-QAOA scheme introduced in
Ref.~\cite{zhukov2025grover}, where the circuit parameters are fixed rather than
optimized. In that work, the authors considered the choice
\begin{equation} \label{eq:const_angles}
    \beta_k = \frac{\pi}{2},
    \qquad
    \gamma_k = -\frac{\pi}{E_{\rm min}^{\rm est}},
\end{equation}
for all layers $k$. This parametrization closely parallels Grover's search
algorithm: the choice $\beta_k = \pi/2$ plays the role of the Grover diffusion
operator, while $\gamma_k = -\pi/E_{\rm min}^{\rm est}$ induces a phase shift of
approximately $-1$ on the target (minimum-energy) state. We refer to this constant-angle variant as GM-QAOA(c).

To study analytically optimized angle selection, we consider ensembles of
SK problem instances with interaction orders
$D = 2, 3, 4$ and system sizes $n = 5, \ldots, 14$. For each pair $(D,n)$, we
generate 100 random instances.
We focus on SK problems because their coupling coefficients are drawn from a
normal distribution, which naturally supports the Gaussian approximation of the
energy spectrum underlying our extreme-value-theory-based analysis.

\begin{figure}
\centering
\includegraphics[width=\textwidth]{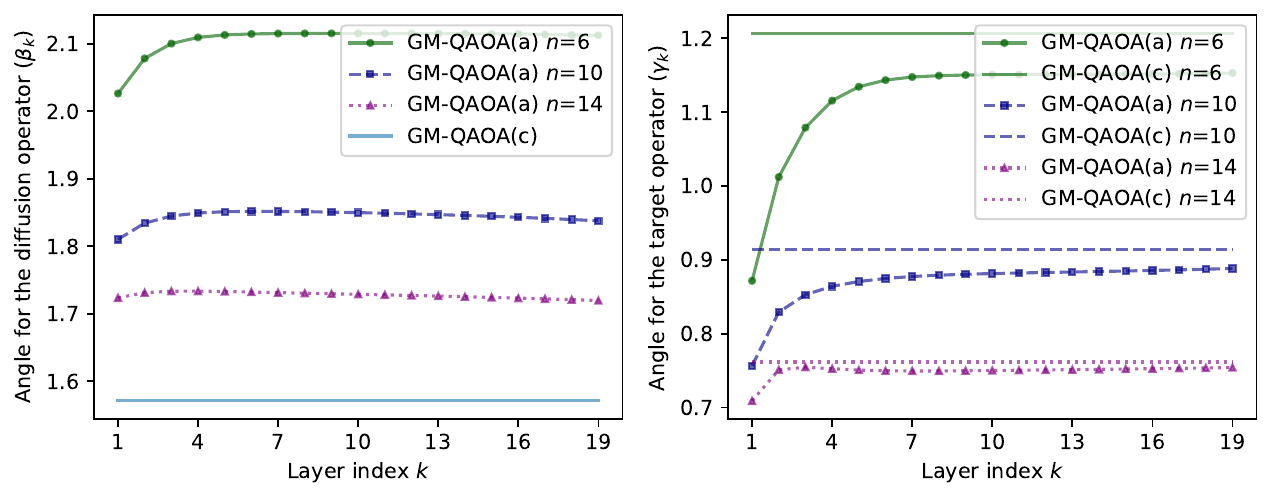}
\caption{Comparison of analytically optimized parameters $\beta_k$ (left panel) and $\gamma_k$ (right panel) as functions of the layer index $k$ for GM-QAOA(a), contrasted with the constant-angle variant GM-QAOA(c) introduced in Ref.~\cite{zhukov2025grover}. The data are averaged over 100 random SK instances with
HUBO order $D=2$ (corresponding to the Ising case) for each system size $n$.
}
\label{fig:comp_angles_vniia_analyt}
\end{figure}

In Fig.~\ref{fig:comp_angles_vniia_analyt}, we compare the optimized angle values
$\{\gamma_k,\beta_k\}$ obtained from the analytical optimization procedure with
the constant-angle choice defined in Eq.~\eqref{eq:const_angles} for $D=2$. The comparison
reveals several notable features.
First, the optimal mixing angle $\beta_k$ in the analytically optimized scheme is
not constant but exhibits a clear dependence on the system size $n$. Its asymptotic value at large $n$ deviates significantly from $\pi/2$, approaching a substantially larger value. Second, while the asymptotic behavior of the phase angles $\gamma_k$ shows reasonable agreement between the two approaches, the key difference lies in their layerwise structure.
In contrast to the constant-parameter baseline, the analytically optimized angles are strongly layer dependent: both $\beta_k$ and $\gamma_k$ undergo pronounced
adjustments in the initial layers of the circuit before converging to steady-state
values at larger depths. This nontrivial parameter scheduling in the early stages
of the algorithm plays a crucial role in enhancing performance.

\begin{figure}
    \centering
    \includegraphics[width=\textwidth]{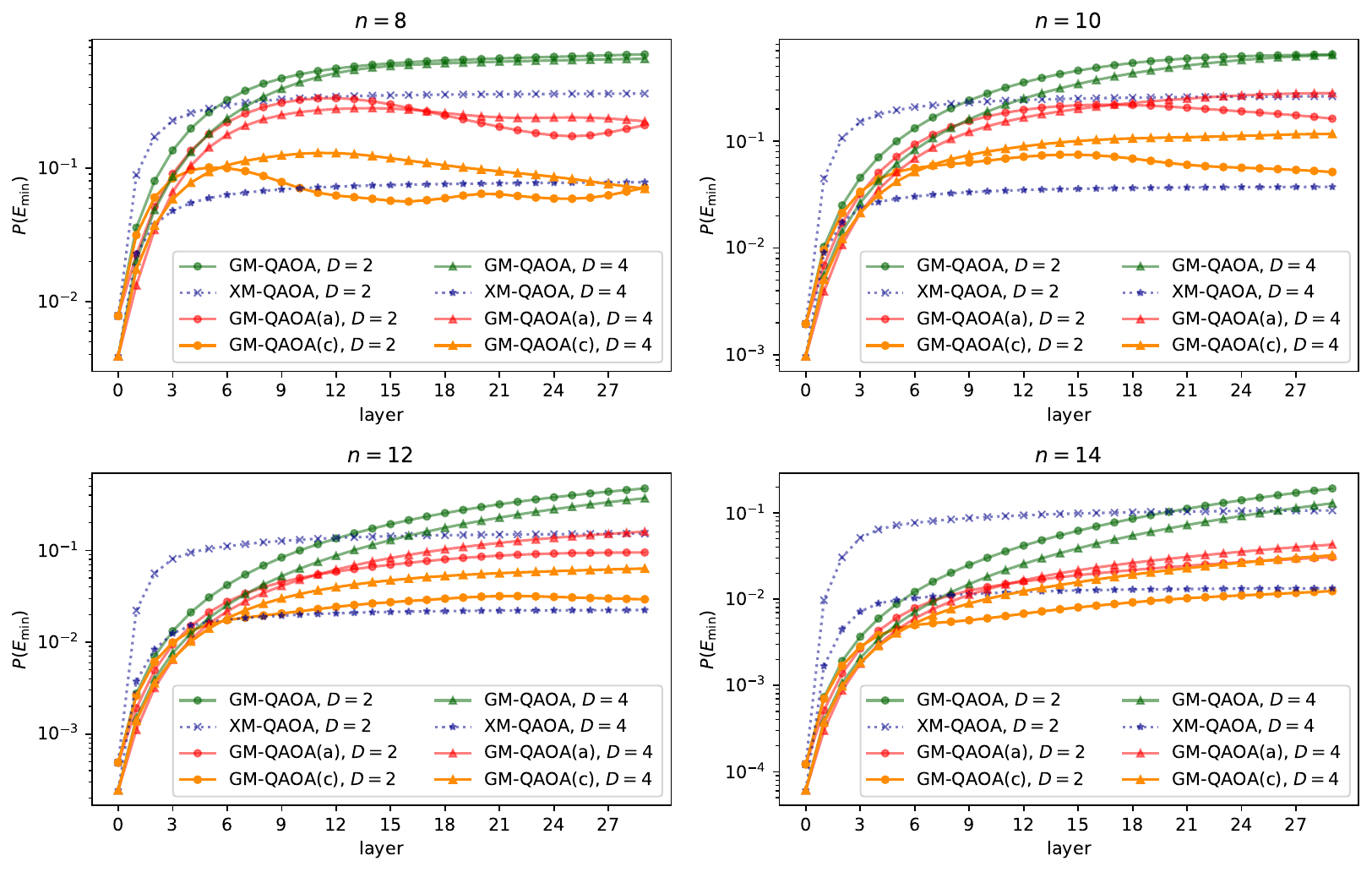}
    \caption{
    Behavior of the success probability $P(E_{\rm min})$ as a function of the number
    of layers for four methods: layerwise optimized GM-QAOA, XM-QAOA, and two
    analytical approaches—one with parameter optimization [GM-QAOA(a)] and one without
    [GM-QAOA(c)] --- for various problem sizes $n$ and HUBO orders $D$.
    }
    \label{fig:comp_prob_algo_analyt_sk}
\end{figure}

Next, we compare the performance of XM- and GM-QAOA with layerwise optimization, introduced earlier, against two analytical approaches --- one with parameter optimization and one without. Illustrative results are shown in Fig.~\ref{fig:comp_prob_algo_analyt_sk}.

Several key conclusions can be drawn from these results. First, the analytically optimized approach [GM-QAOA(a)] consistently outperforms its nonoptimized counterpart [GM-QAOA(c)], highlighting the importance of classical preoptimization of the mixing angles $\beta_k$.
Second, as the locality of the cost Hamiltonian increases the performance of GM-QAOA(a) improves significantly, approaching that of the layerwise optimized GM-QAOA while requiring substantially fewer quantum resources. Notably, for $D=4$ we observe that beyond a
certain circuit depth the analytically optimized variant GM-QAOA(a) begins to outperform XM-QAOA.  We also note that for layerwise optimized GM-QAOA, increasing the locality from $D=2$ to $4$ leads to a degradation of the success probability $P(E_{\rm min})$. In
contrast, for the analytically motivated approaches this trend is reversed, with higher-order interactions yielding improved performance. This behavior suggests that analytical angle selection becomes increasingly advantageous as the locality of the cost Hamiltonian grows.

\begin{figure}
    \centering
    \includegraphics[width=0.6 \textwidth]{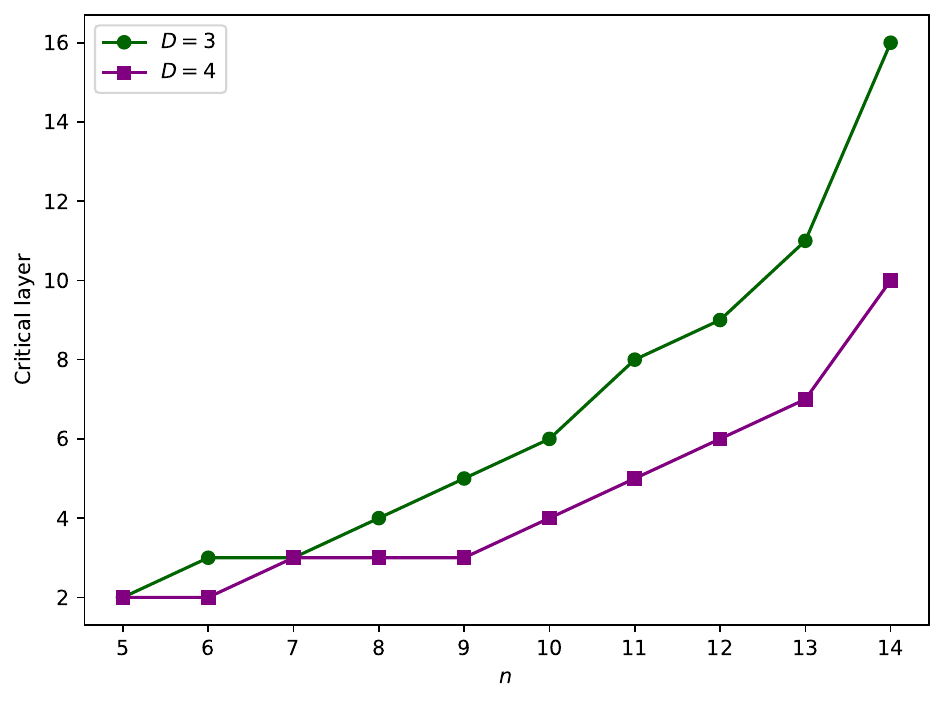}
    \caption{
    Minimum circuit depth at which GM-QAOA(a) outperforms XM-QAOA in terms of the success probability $P(E_{\rm min})$ as a function of the number of qubits $n$ for $D=3,4$.}
    \label{fig:phase_graph_analytics}
\end{figure}

To conclude our analysis, we investigate the behavior of the crossover point,
defined as the critical circuit depth (layer index) beyond which GM-QAOA begins to outperform XM-QAOA in terms of the success probability $P(E_{\rm min})$, averaged over problem instances. Figure~\ref{fig:phase_graph_analytics} shows the critical circuit depths at which the analytically optimized GM-QAOA(a) surpasses the performance plateau of XM-QAOA. For the case of pairwise interactions ($D=2$), no such crossover is observed within the considered range of system sizes, and XM-QAOA consistently outperforms GM-QAOA(a) for all $n$.
The observed behavior exhibits the same qualitative characteristics as the \BLUE{layerwise} optimized GM-QAOA implementation. In particular, the critical depth increases
monotonically with the problem size $n$ while displaying an inverse dependence on
the interaction order $D$ of the cost Hamiltonian.

\section{Conclusions}\label{sec:Conclusion}\label{sec:concl}

In this work, we have systematically investigated the application of GM-QAOA to
HUBO problems. Our results demonstrate that the GM-QAOA framework, owing to its
global mixing mechanism, offers a distinct advantage over XM-QAOA for optimization
problems involving high-order interactions. In particular, we have shown that
while XM-QAOA exhibits an early performance plateau at shallow circuit depths,
GM-QAOA displays a monotonic improvement with increasing depth and ultimately
surpasses XM-QAOA beyond a critical layer. This advantage becomes increasingly
pronounced as the locality $D$ of the cost Hamiltonian grows, highlighting the
robustness of GM-QAOA against the complexity induced by multi-spin correlations.

\BLUE{We note that the layerwise optimization employed in this work is a practical heuristic and may yield suboptimal performance compared to global optimization, particularly for the transverse-field mixer. Accordingly, our comparison is strictly valid within this optimization framework. Within this setting, however, the observed behavior suggests that GM-QAOA is less sensitive to suboptimal parameter choices, which may be advantageous for near-term implementations with limited optimization resources. Assessing whether this advantage persists under globally optimized parameters remains an important direction for future work.}

A key contribution of this study is the development of an analytical model for
GM-QAOA dynamics with variable mixing angles.
Building on this model, we introduced a resource-efficient parameter selection
strategy based on classical preoptimization. We demonstrated that the resulting
analytically optimized variant, GM-QAOA(a), can achieve performance close to that
of the \BLUE{layerwise} optimized GM-QAOA, while substantially reducing the required quantum
resources. This establishes GM-QAOA(a) as a practical and scalable approach for
high-order optimization problems.

Several promising directions for future research naturally follow from our results. First, extending the present analytical framework to alternative mixer Hamiltonians and broader classes of optimization problems, including constrained settings, could further enhance the versatility of QAOA-based approaches. Second, an experimental realization of GM-QAOA for HUBO problems on qudit-based quantum processors represents an important next step toward practical validation, particularly in view of the more compact and native encodings enabled by multilevel systems. Finally, a promising avenue for future study is the development of fixed-point strategies that simultaneously scale the system size and circuit depth, while employing low-dimensional parametrizations of the variational parameters~\cite{chernyavskiy2025improving}. Taken together, these directions suggest that the proposed framework contributes to advancing quantum optimization toward practical relevance for complex, large-scale problems.

\section*{acknowledgment}
{This research was supported by the Priority 2030 program at the National University of Science and Technology ``MISIS'' under Project No. K1-2022-027.}

\section*{Data availability}
\BLUE{
The datasets generated and analyzed during this study, including ensembles of random Hamiltonians and optimized QAOA parameters are publicly available at~\cite{supplementary_data}.
}

\bibliographystyle{apsrev4-2}
\bibliography{references.bib}

@article{rosenberg1975reduction,
  title={Reduction of bivalent maximization to the quadratic case},
  author={Rosenberg, Ivo G},
  journal={Cahiers du Centre d'Études de Recherche Opérationnelle},
  volume={17},
  number={1},
  pages={71--74},
  year={1975}
}

@article{boros2014quadratization,
  title={On quadratization of pseudo-Boolean functions},
  author={Boros, Endre and Gruber, Aritanan},
  journal={arXiv preprint arXiv:1404.6538},
  year={2014}
}

@article{dattani2019quadratization,
  title={Quadratization in discrete optimization and quantum mechanics},
  author={Dattani, Nike},
  journal={arXiv preprint arXiv:1901.04405},
  year={2019}
}

@article{wei2014detecting,
  title={Detecting epistasis in human complex traits},
  author={Wei, Wen-Hua and Hemani, Gibran and Haley, Chris S},
  journal={Nature Reviews Genetics},
  volume={15},
  number={11},
  pages={722--733},
  year={2014},
  publisher={Nature Publishing Group UK London}
}

@article{klein2008large,
  title={Large-scale molecular dynamics simulations of self-assembling systems},
  author={Klein, Michael L and Shinoda, Wataru},
  journal={science},
  volume={321},
  number={5890},
  pages={798--800},
  year={2008},
  publisher={American Association for the Advancement of Science}
}

@inproceedings{sejnowski1986higher,
  title={Higher-order Boltzmann machines},
  author={Sejnowski, Terrence J and others},
  booktitle={AIP Conference Proceedings},
  volume={151},
  number={1},
  pages={398--403},
  year={1986},
  organization={American Institute of Physics}
}

@article{kolda2009tensor,
  title={Tensor decompositions and applications},
  author={Kolda, Tamara G and Bader, Brett W},
  journal={SIAM review},
  volume={51},
  number={3},
  pages={455--500},
  year={2009},
  publisher={SIAM}
}

@article{chang2011libsvm,
  title={LIBSVM: A library for support vector machines},
  author={Chang, Chih-Chung and Lin, Chih-Jen},
  journal={ACM transactions on intelligent systems and technology (TIST)},
  volume={2},
  number={3},
  pages={1--27},
  year={2011},
  publisher={Acm New York, NY, USA}
}

@article{ribeiro2008hybrid,
  title={A hybrid heuristic for a multi-objective real-life car sequencing problem with painting and assembly line constraints},
  author={Ribeiro, Celso C and Aloise, Daniel and Noronha, Thiago F and Rocha, Caroline and Urrutia, Sebasti{\'a}n},
  journal={European Journal of Operational Research},
  volume={191},
  number={3},
  pages={981--992},
  year={2008},
  publisher={Elsevier}
}

@article{bierwirth1999production,
  title={Production scheduling and rescheduling with genetic algorithms},
  author={Bierwirth, Christian and Mattfeld, Dirk C},
  journal={Evolutionary computation},
  volume={7},
  number={1},
  pages={1--17},
  year={1999},
  publisher={MIT Press One Rogers Street, Cambridge, MA 02142-1209, USA journals-info~…}
}

@article{farhi2014quantum,
  title={A quantum approximate optimization algorithm},
  author={Farhi, Edward and Goldstone, Jeffrey and Gutmann, Sam},
  journal={arXiv preprint arXiv:1411.4028},
  year={2014}
}

@inproceedings{grover1996fast,
  title={A fast quantum mechanical algorithm for database search},
  author={Grover, Lov K},
  booktitle={Proceedings of the twenty-eighth annual ACM symposium on Theory of computing},
  pages={212--219},
  year={1996}
}

@article{kempe2006complexity,
  title={The complexity of the local Hamiltonian problem},
  author={Kempe, Julia and Kitaev, Alexei and Regev, Oded},
  journal={Siam journal on computing},
  volume={35},
  number={5},
  pages={1070--1097},
  year={2006},
  publisher={SIAM}
}

@article{tsvelikhovskiy2025provable,
  title={Provable avoidance of barren plateaus for the Quantum Approximate Optimization Algorithm with Grover mixers},
  author={Tsvelikhovskiy, Boris and Nuyten, Matthew and Bakalov, Bojko N},
  journal={arXiv preprint arXiv:2509.10424},
  year={2025}
}

@article{zhukov2025grover,
  title={Grover's search meets Ising models: a quantum algorithm for finding low-energy states},
  author={Zhukov, AA and Lebedev, AV and Pogosov, WV},
  journal={Computer Physics Communications},
  pages={109627},
  year={2025},
  publisher={Elsevier}
}

@inproceedings{bartschi2020grover,
 title={Grover mixers for QAOA: Shifting complexity from mixer design to state preparation},
 author={B{\"a}rtschi, Andreas and Eidenbenz, Stephan},
 booktitle={2020 IEEE International Conference on Quantum Computing and Engineering (QCE)},
 pages={72--82},
 year={2020},
 organization={IEEE}
}

@article{bridi2024analytical,
 title={Analytical results for the quantum alternating operator ansatz with Grover mixer},
 author={Bridi, Guilherme Adamatti and Marquezino, Franklin de Lima},
 journal={Physical Review A},
 volume={110},
 number={5},
 pages={052409},
 year={2024},
 publisher={APS}
}

@article{xie2025performance,
 title={Performance upper bound of a Grover-mixer quantum alternating operator ansatz},
 author={Xie, Ningyi and Xu, Jiahua and Chen, Tiejin and Lee, Xinwei and Saito, Yoshiyuki and Asai, Nobuyoshi and Cai, Dongsheng},
 journal={Physical Review A},
 volume={111},
 number={1},
 pages={012401},
 year={2025},
 publisher={APS}
}

@inproceedings{golden2021threshold,
 title={Threshold-based quantum optimization},
 author={Golden, John and B{\"a}rtschi, Andreas and O’Malley, Daniel and Eidenbenz, Stephan},
 booktitle={2021 IEEE International Conference on Quantum Computing and Engineering (QCE)},
 pages={137--147},
 year={2021},
 organization={IEEE}
}

@article{benchasattabuse2023lower,
 title={Lower bounds on number of QAOA rounds required for guaranteed approximation ratios},
 author={Benchasattabuse, Naphan and B{\"a}rtschi, Andreas and Garc{\'\i}a-Pintos, Luis Pedro and Golden, John and Lemons, Nathan and Eidenbenz, Stephan},
 journal={arXiv preprint arXiv:2308.15442},
 year={2023}
}

@article{pelofske2025biased,
 title={Biased degenerate ground-state sampling of small Ising models with converged quantum approximate optimization algorithm},
 author={Pelofske, Elijah},
 journal={Physical Review E},
 volume={111},
 number={5},
 pages={054103},
 year={2025},
 publisher={APS}
}

@article{ng2024analytical,
 title={Analytical expressions for the quantum approximate optimization algorithm and its variants},
 author={Ng, Truman Yu and Koh, Jin Ming and Koh, Dax Enshan},
 journal={arXiv preprint arXiv:2411.09745},
 year={2024}
}

@article{jenkinson1955frequency,
 title={The frequency distribution of the annual maximum (or minimum) values of meteorological elements},
 author={Jenkinson, Arthur F},
 journal={Quarterly Journal of the Royal meteorological society},
 volume={81},
 number={348},
 pages={158--171},
 year={1955},
 publisher={Wiley}
}

@article{pickands1975statistical,
 title={Statistical inference using extreme order statistics},
 author={Pickands III, James},
 journal={the Annals of Statistics},
 pages={119--131},
 year={1975},
 publisher={JSTOR}
}

@book{coles2001introduction,
 title={An introduction to statistical modeling of extreme values},
 author={Coles, Stuart and Bawa, Joanna and Trenner, Lesley and Dorazio, Pat},
 volume={208},
 year={2001},
 publisher={Springer}
}

@book{beirlant2006statistics,
 title={Statistics of extremes: theory and applications},
 author={Beirlant, Jan and Goegebeur, Yuri and Segers, Johan and Teugels, Jozef L},
 year={2006},
 publisher={John Wiley \& Sons}
}

@book{haan2006extreme,
 title={Extreme value theory: an introduction},
 author={Haan, Laurens and Ferreira, Ana},
 volume={3},
 year={2006},
 publisher={Springer}
}

@inproceedings{fisher1928limiting,
  title={Limiting forms of the frequency distribution of the largest or smallest member of a sample},
  author={Fisher, Ronald Aylmer and Tippett, Leonard Henry Caleb},
  booktitle={Mathematical proceedings of the Cambridge philosophical society},
  volume={24},
  number={2},
  pages={180--190},
  year={1928},
  organization={Cambridge University Press}
}

@article{gnedenko1943distribution,
  title={Sur la distribution limite du terme maximum d'une serie aleatoire},
  author={Gnedenko, Boris},
  journal={Annals of mathematics},
  volume={44},
  number={3},
  pages={423--453},
  year={1943},
  publisher={JSTOR}
}

@article{kiktenko2025colloquium,
  title={Colloquium: Qudits for decomposing multiqubit gates and realizing quantum algorithms},
  author={Kiktenko, Evgeniy O and Nikolaeva, Anastasiia S and Fedorov, Aleksey K},
  journal={Reviews of Modern Physics},
  volume={97},
  number={2},
  pages={021003},
  year={2025},
  publisher={APS}
}

@article{nikolaeva2024efficient,
  title={Efficient realization of quantum algorithms with qudits},
  author={Nikolaeva, Anastasiia S and Kiktenko, Evgeniy O and Fedorov, Aleksey K},
  journal={EPJ Quantum Technology},
  volume={11},
  number={1},
  pages={1--25},
  year={2024},
  publisher={Springer}
}

@article{nikolaeva2025scalable,
  title={Scalable improvement of the generalized Toffoli gate realization using trapped-ion-based qutrits},
  author={Nikolaeva, Anastasiia S and Zalivako, Ilia V and Borisenko, Alexander S and Semenin, Nikita V and Galstyan, Kristina P and Korolkov, Andrey E and Kiktenko, Evgeniy O and Khabarova, Ksenia Yu and Semerikov, Ilya A and Fedorov, Aleksey K and others},
  journal={Physical Review Letters},
  volume={135},
  number={6},
  pages={060601},
  year={2025},
  publisher={APS}
}

@article{chu2023scalable,
  title={Scalable algorithm simplification using quantum AND logic},
  author={Chu, Ji and He, Xiaoyu and Zhou, Yuxuan and Yuan, Jiahao and Zhang, Libo and Guo, Qihao and Hai, Yongju and Han, Zhikun and Hu, Chang-Kang and Huang, Wenhui and others},
  journal={Nature physics},
  volume={19},
  number={1},
  pages={126--131},
  year={2023},
  publisher={Nature Publishing Group UK London}
}

@article{kiktenko2020scalable,
  title={Scalable quantum computing with qudits on a graph},
  author={Kiktenko, Evgeniy O and Nikolaeva, Anastasiia S and Xu, Peng and Shlyapnikov, Georgy V and Fedorov, Arkady K},
  journal={Physical Review A},
  volume={101},
  number={2},
  pages={022304},
  year={2020},
  publisher={APS}
}

@article{nikolaeva2022decomposing,
  title={Decomposing the generalized Toffoli gate with qutrits},
  author={Nikolaeva, Anastasiia S and Kiktenko, Evgeniy O and Fedorov, Aleksey K},
  journal={Physical review A},
  volume={105},
  number={3},
  pages={032621},
  year={2022},
  publisher={APS}
}

@article{shaydulin2024evidence,
  title={Evidence of scaling advantage for the quantum approximate optimization algorithm on a classically intractable problem},
  author={Shaydulin, Ruslan and Li, Changhao and Chakrabarti, Shouvanik and DeCross, Matthew and Herman, Dylan and Kumar, Niraj and Larson, Jeffrey and Lykov, Danylo and Minssen, Pierre and Sun, Yue and others},
  journal={Science Advances},
  volume={10},
  number={22},
  pages={eadm6761},
  year={2024},
  publisher={American Association for the Advancement of Science}
}

@article{fedorov2022quantum,
  title={Quantum computing at the quantum advantage threshold: a down-to-business review},
  author={Fedorov, Arkady K and Gisin, Nicolas and Beloussov, Serguei M and Lvovsky, Alexander I},
  journal={arXiv preprint arXiv:2203.17181},
  year={2022}
}

@article{semenov2025technique,
 title={Technique for Transforming Discrete Optimization Problems into QUBO Form},
 author={Semenov, AM and Usmanov, SR and Fedorov, AK},
 journal={Problems of Information Transmission},
 volume={61},
 number={2},
 pages={110--142},
 year={2025},
 publisher={Springer}
}

@article{chermoshentsev2021polynomial,
 title={Polynomial unconstrained binary optimisation inspired by optical simulation},
 author={Chermoshentsev, Dmitry A and Malyshev, Aleksei O and Esencan, Mert and Tiunov, Egor S and Mendoza, Douglas and Aspuru-Guzik, Al{\'a}n and Fedorov, Aleksey K and Lvovsky, Alexander I},
 journal={arXiv preprint arXiv:2106.13167},
 year={2021}
}

@article{farhi2016quantum,
 title={Quantum supremacy through the quantum approximate optimization algorithm},
 author={Farhi, Edward and Harrow, Aram W},
 journal={arXiv preprint arXiv:1602.07674},
 year={2016}
}

@article{pagano2020quantum,
 title={Quantum approximate optimization of the long-range Ising model with a trapped-ion quantum simulator},
 author={Pagano, Guido and Bapat, Aniruddha and Becker, Patrick and Collins, Katherine S and De, Arinjoy and Hess, Paul W and Kaplan, Harvey B and Kyprianidis, Antonis and Tan, Wen Lin and Baldwin, Christopher and others},
 journal={Proceedings of the National Academy of Sciences},
 volume={117},
 number={41},
 pages={25396--25401},
 year={2020},
 publisher={National Academy of Sciences}
}

@article{harrigan2021quantum,
  title={Quantum approximate optimization of non-planar graph problems on a planar superconducting processor},
  author={Harrigan, Matthew P and Sung, Kevin J and Neeley, Matthew and Satzinger, Kevin J and Arute, Frank and Arya, Kunal and Atalaya, Juan and Bardin, Joseph C and Barends, Rami and Boixo, Sergio and others},
  journal={Nature Physics},
  volume={17},
  number={3},
  pages={332--336},
  year={2021},
  publisher={Nature Publishing Group UK London}
}

@article{zhou2020quantum,
  title={Quantum approximate optimization algorithm: Performance, mechanism, and implementation on near-term devices},
  author={Zhou, Leo and Wang, Sheng-Tao and Choi, Soonwon and Pichler, Hannes and Lukin, Mikhail D},
  journal={Physical Review X},
  volume={10},
  number={2},
  pages={021067},
  year={2020},
  publisher={APS}
}

@article{biamonte2017quantum,
  title={Quantum machine learning},
  author={Biamonte, Jacob and Wittek, Peter and Pancotti, Nicola and Rebentrost, Patrick and Wiebe, Nathan and Lloyd, Seth},
  journal={Nature},
  volume={549},
  number={7671},
  pages={195--202},
  year={2017},
  publisher={Nature Publishing Group UK London}
}

@article{mcardle2020quantum,
  title={Quantum computational chemistry},
  author={McArdle, Sam and Endo, Suguru and Aspuru-Guzik, Al{\'a}n and Benjamin, Simon C and Yuan, Xiao},
  journal={Reviews of Modern Physics},
  volume={92},
  number={1},
  pages={015003},
  year={2020},
  publisher={APS}
}

@article{xu2021variational,
  title={Variational algorithms for linear algebra},
  author={Xu, Xiaosi and Sun, Jinzhao and Endo, Suguru and Li, Ying and Benjamin, Simon C and Yuan, Xiao},
  journal={Science Bulletin},
  volume={66},
  number={21},
  pages={2181--2188},
  year={2021},
  publisher={Elsevier}
}

@article{yuan2019theory,
  title={Theory of variational quantum simulation},
  author={Yuan, Xiao and Endo, Suguru and Zhao, Qi and Li, Ying and Benjamin, Simon C},
  journal={Quantum},
  volume={3},
  pages={191},
  year={2019},
  publisher={Verein zur F{\"o}rderung des Open Access Publizierens in den Quantenwissenschaften}
}

@article{bharti2022noisy,
  title={Noisy intermediate-scale quantum algorithms},
  author={Bharti, Kishor and Cervera-Lierta, Alba and Kyaw, Thi Ha and Haug, Tobias and Alperin-Lea, Sumner and Anand, Abhinav and Degroote, Matthias and Heimonen, Hermanni and Kottmann, Jakob S and Menke, Tim and others},
  journal={Reviews of Modern Physics},
  volume={94},
  number={1},
  pages={015004},
  year={2022},
  publisher={APS}
}

@article{cerezo2021variational,
  title={Variational quantum algorithms},
  author={Cerezo, Marco and Arrasmith, Andrew and Babbush, Ryan and Benjamin, Simon C and Endo, Suguru and Fujii, Keisuke and McClean, Jarrod R and Mitarai, Kosuke and Yuan, Xiao and Cincio, Lukasz and others},
  journal={Nature Reviews Physics},
  volume={3},
  number={9},
  pages={625--644},
  year={2021},
  publisher={Nature Publishing Group UK London}
}

@article{chernyavskiy2025improving,
  title={Improving QAOA to find approximate QUBO solutions in O (1) shots},
  author={Chernyavskiy, A Yu and Kulikov, DA and Bantysh, BI and Bogdanov, Yu I and Fedorov, AK and Kiktenko, EO},
  journal={arXiv preprint arXiv:2509.19035},
  year={2025}
}

@inproceedings{golden2023quantum,
 title={The quantum alternating operator ansatz for satisfiability problems},
 author={Golden, John and B{\"a}rtschi, Andreas and O'Malley, Daniel and Eidenbenz, Stephan},
 booktitle={2023 IEEE International Conference on Quantum Computing and Engineering (QCE)},
 volume={1},
 pages={307--312},
 year={2023},
 organization={IEEE}
}

@inproceedings{golden2023numerical,
 title={Numerical evidence for exponential speed-up of QAOA over unstructured search for approximate constrained optimization},
 author={Golden, John and B{\"a}rtschi, Andreas and O'Malley, Daniel and Eidenbenz, Stephan},
 booktitle={2023 IEEE International Conference on Quantum Computing and Engineering (QCE)},
 volume={1},
 pages={496--505},
 year={2023},
 organization={IEEE}
}

@article{sundar2019quantum,
 title={A quantum algorithm to count weighted ground states of classical spin Hamiltonians},
 author={Sundar, Bhuvanesh and Paredes, Roger and Damanik, David T and Duenas-Osorio, Leonardo and Hazzard, Kaden RA},
 journal={arXiv preprint arXiv:1908.01745},
 year={2019}
}

@article{zhu2023multi,
 title={Multi-round QAOA and advanced mixers on a trapped-ion quantum computer},
 author={Zhu, Yingyue and Zhang, Zewen and Sundar, Bhuvanesh and Green, Alaina M and Huerta Alderete, C and Nguyen, Nhung H and Hazzard, Kaden RA and Linke, Norbert M},
 journal={Quantum Science \& Technology},
 volume={8},
 number={1},
 pages={015007},
 year={2023},
 publisher={IOP Publishing}
}

@article{zhang2025grover,
 title={Grover-QAOA for 3-SAT: quadratic speedup, fair-sampling, and parameter clustering},
 author={Zhang, Zewen and Paredes, Roger and Sundar, Bhuvanesh and Quiroga, David and Kyrillidis, Anastasios and Duenas-Osorio, Leonardo and Pagano, Guido and Hazzard, Kaden RA},
 journal={Quantum Science and Technology},
 volume={10},
 number={1},
 pages={015022},
 year={2025},
 publisher={IOP Publishing}
}

@dataset{supplementary_data,
  author       = {Kiktenko, Evgeniy O and Krendeleva, Elizaveta V. and Fedorov, Aleksey K},
  title        = {Supplementary data for "Applying GM-QAOA to HUBO problems"},
  year         = {2026},
  publisher    = {Zenodo},
  doi          = {10.5281/zenodo.19372925},
  url          = {https://doi.org/10.5281/zenodo.19372925}
}

@inproceedings{uotila2025higher,
  title={Higher-order portfolio optimization with quantum approximate optimization algorithm},
  author={Uotila, Valter and Ripatti, Julia and Zhao, Bo},
  booktitle={2025 IEEE International Conference on Quantum Computing and Engineering (QCE)},
  volume={1},
  pages={01--12},
  year={2025},
  organization={IEEE}
}

@article{ikeuchi2025evaluating,
  title={Evaluating the Performance of Direct Higher-Order Formulations in Combinatorial Optimization Problems},
  author={Ikeuchi, Kazuki and Matsuda, Yoshiki and Tanaka, Shu},
  journal={arXiv preprint arXiv:2510.24237},
  year={2025}
}

@article{akshay2020reachability,
  title={Reachability deficits in quantum approximate optimization},
  author={Akshay, Vishwanathan and Philathong, Hariphan and Morales, Mauro ES and Biamonte, Jacob D},
  journal={Physical review letters},
  volume={124},
  number={9},
  pages={090504},
  year={2020},
  publisher={APS}
}

\end{document}